# Why do hot and cold water sound different when poured?

Xiaotian Bi[a], Dike Su[b], Qianyun Zhou[a]

a. *Argon-Hydrogen Studio, Beijing, China*

b. *Communication University of China, Beijing, China*

Correspondence should be addressed to Xiaotian Bi (bxt10@foxmail.com)

**Abstract** Empirical studies have demonstrated that humans possess the remarkable capacity to distinguish whether a glass of water is hot or cold solely by the sound of pouring it. However, the underlying physical mechanisms governing the disparities in the acoustic signatures of hot versus cold water remain to be deciphered. In this paper, we conducted a series of experiments to extract the intrinsic features of pouring sounds at contrasting temperatures. The results of our spectral analysis revealed that the sound of pouring hot water exhibited more pronounced low-frequency components and diminished high-frequency components relative to cold water. High-speed photographic evidence elucidated that pouring hot water could generate larger air bubbles in greater abundance. We conjecture that the Minnaert resonance arising from these larger entrained bubbles in hot water produces a lower-frequency acoustic signature, thereby constituting the foundational mechanistic explanation for the auditory distinction between pouring hot and cold water.

**Keywords** Fluid acoustics, Minnaert resonance, water sound, hot and cold water, bubble entrainment

## 1. Introduction

The soundscapes that pervade our daily lives are frequently replete with the acoustic signatures of water, encompassing the melodic babbling of brooks, the thunderous roar of waterfalls, and the crashing resonance of oceanic waves. One such occurrence among these ubiquitous aquatic sonances, often overlooked in its simplicity, is the sound produced by pouring water. The simple act of pouring can excite the vibrations of multiple media, including the water itself, its containing vessel, the air, and underwater bubbles, thereby generating complex acoustic emissions governed by a combination of diverse physical factors (Moss et al., 2010). In 2013, Velasco et al. drew attention to the intriguing phenomenon wherein humans can accurately hear the differences between the sounds produced by hot and cold water (Velasco et al., 2013). While this auditory perception of water temperature has garnered significant interest in psychology and sensory science (Spence, 2020, 2021; Agrawal and Schachner, 2023), it has also inspired a quest to elucidate the physical mechanisms by which temperature influences the acoustics of fluids.

In 2017, science communicators Tom Scott and Steve Mould proposed that the difference in sound between cold and hot water stemmed from differences in viscosity

(*You Can Hear The Difference Between Hot and Cold Water*, 2017). However, Steve Mould acknowledged that the intricate fluid dynamics render an accurate explanation challenging. Subsequently, Peng and Reiss identified the three sound sources of pouring: air resonance, vibrations of the container and water, and bubble sounds (Peng and Reiss, 2018). Through systematic analysis of the frequency spectrum, they concluded that the intensities of container and water vibrations diminish more rapidly than those of air resonance, potentially underlying the audible difference in pouring sounds. Their work is virtually the only acoustic study of this problem in academia, but it did not build a bridge between physics and acoustics. More recently, Rohlin and Thulin proposed a signal analysis and machine learning algorithm to predict water temperature from audio samples of pouring, offering an intriguing potential applications of this phenomenon (Rohlin and Thulin, 2022). So far, the physics underpinning the auditory distinction between pouring hot and cold water remains to be deciphered.

In order to explore the magic behind this problem, it is instructive to review the fundamental principles governing how water generates sound. A century ago, Sir Bragg elucidated in his interesting book *The World of Sound* that the sounds of water mainly originate from the formation of bubbles within the fluid (Bragg, 1921). Experience in the outdoors, such as the absence of audible sounds from smoothly flowing creeks without bubbles, corroborate this principle. Although water can also produce sounds through other mechanisms (Franz, 1959), the entrainment of air bubbles is widely regarded as the primary source of water sounds (Leighton, 1994; Strasberg, 1956). A video presentation of this argument can be seen in Supplementary Material 1.

The subtle variations in bubble characteristics contribute to the rich diversity of aquatic acoustics that humans have honed the ability to discriminate (Doel, 2005). The fact that you can discern the sounds of boiling water, pouring into a glass, showers, streams, and waterfalls is predicated upon the distinct bubble signatures in each scenario (Guyot et al., 2017). In the previous work of Peng and Reiss, they attributed the key acoustic features of pouring hot and cold water to air column resonance and vibrations of the container and water. However, we believe that their assertion overlooks the significant contribution of bubble sounds to the auditory perception.

Intuitive observations suggest that the sound of pouring cold water exhibits a crisper and more continuous character, while the sound of pouring hot water is perceived as duller and splashier. What exactly characterizes them in terms of frequency? Why is there such a difference? The phenomenon under investigation lies at the interdisciplinary nexus of acoustics and fluid dynamics. The objective of this paper is to delve into these two subjects, with a focus on the generation of distinct bubble populations at different temperatures. First, the three primary sound components of pouring water and their characteristic features will be clarified. Second, the frequency spectra of the sounds generated by pouring hot and cold water will be analyzed to elucidate the essential differences between their acoustic features. Third, the effect of bubble size distributions on the frequency of the bubble sound will be examined to explain the observed disparities in the spectra of water sounds. In short, we hope to provide a preliminary exploration of the physical principles behind this interesting auditory phenomenon.

It is necessary to mention that the first author of this paper, despite being a former chemical engineering student, is not currently an active scientific researcher but rather an enthusiast with a keen interest in science. Since he was lucky enough to graduate last year, he now has neither the resources of a lab nor the guidance of a faculty member in a related field. Most of the work in this paper originates from his monkeying around. Consequently, this paper presents only a superficial experimental exploration and rudimentary theoretical analysis of the subject matter. We are keenly aware of the low academic quality of this article, and openly acknowledge the potential for errors or inaccuracies within its content. Nonetheless, for the majority of individuals harboring much curiosity regarding the acoustic phenomena associated with pouring hot and cold water, we anticipate that this paper may provide engaging insights and the joy of observing life.

## 2. What makes up the sound of pouring water?

As Peng and Reiss points out, the sound of pouring water into an empty vessel consists of three parts: the resonance of the air, the vibration of the vessel and the water inside, and the sound of bubbles. In the following we will briefly describe the physical model of each sound source.

### 2.1 Resonance of the air

When filling a vessel with water, one can observe the presence of an audible tone exhibiting a rise in frequency. This rising tone becomes particularly noticeable when the vessel possesses a large length-to-diameter ratio, such as a glass bottle. The fundamental resonant frequency of an air column in a tube closed at one end can be calculated as follows (Cabe and Pittenger, 2000).

$$f_{air} = \frac{c}{4(L + 0.62R)}$$

In the above equation, $c$ is the speed of sound, $L$ is the air column length, $R$ is the radius of the vessel.

A cylindrical vessel can also be modelled as a Helmholtz resonator, where the air above the liquid surface resonates when water is poured in (Balachandran et al., 2011; Monteiro et al., 2015). The resonant frequency can be empirically calculated as follows.

$$f_H = \frac{c}{2\pi}\sqrt{\frac{1}{1.7LR}}$$

For both models, the resonant frequency exhibited an inverse relationship with the length of the air column. Therefore, as more water is poured into the vessel, the pitch of the air resonance becomes higher.

A demo video is available in Supplementary Material 2 (A great demonstration experiment for a physics classroom! Although it requires the teacher to have strong blowing skills…).

## 2.2 Vibration of the vessel and the water inside

In the initial moments when the water is poured into the vessel, a crisp tinkling sound can be heard. This acoustic phenomenon arises from the vibrations induced in the vessel itself at its natural resonant frequencies upon impact with the inflowing water (French, 1983).

$$f_0 = \frac{1}{2\pi}\sqrt{\frac{3Y}{5\rho_c}\frac{a}{R^2}}\sqrt{1+\frac{4}{3}(\frac{R}{H})^4}$$

In the above formula, $f_0$ is the vibration frequency of the vessel, $Y$ is the Young's modulus of the vessel, $\rho_c$ is the density of the vessel, $a$ is the thickness of the cylindrical vessel's wall, $R$ is the radius of the vessel, and $H$ is the height of the vessel.

As more water is poured into the glass, the overall mass increases and the frequency of the vibration decreases, conforming to the following relationship (Courtois et al., 2008).

$$f_c = \frac{f_0}{\sqrt{1+\frac{4\rho_l R}{9\rho_c a}(\frac{h}{H})^3}}$$

In the above formula, $f_c$ is the vibration frequency of the vessel and the water inside as a whole, $\rho_l$ is the density of water, $h$ is the height of water. Therefore, as more water is poured into the vessel, the frequency of this vibration becomes lower.

A demo video is available in Supplementary Material 3 (Another great demonstration experiment for a physics classroom!).

## 2.3 Resonance of the bubbles

When pouring water, a free-falling jet stream penetrates the water surface within the receiving vessel at a certain velocity. The Plateau-Rayleigh instability induces rippling on the surface of the jet, which consequently impacts the formation of a large cavity, thereby facilitating the entrainment of bubbles upon entry into the bulk fluid (Boudina et al., 2023).
Immediately upon the closure of the cavity and the formation of the bubble, the bubble surrounded by fluid experiences an increase in pressure attributed to surface tension. This pressure pulsation facilitates the periodic expansion and contraction of the bubble as it interacts with the surrounding water, thereby creating a resonator (Zheng and James, 2009). Then the bubble emits a "bo" or "dong" sound which decays as energy is soon dissipated (Doel, 2005). Figure 1 shows the evolution of an acoustic bubble.

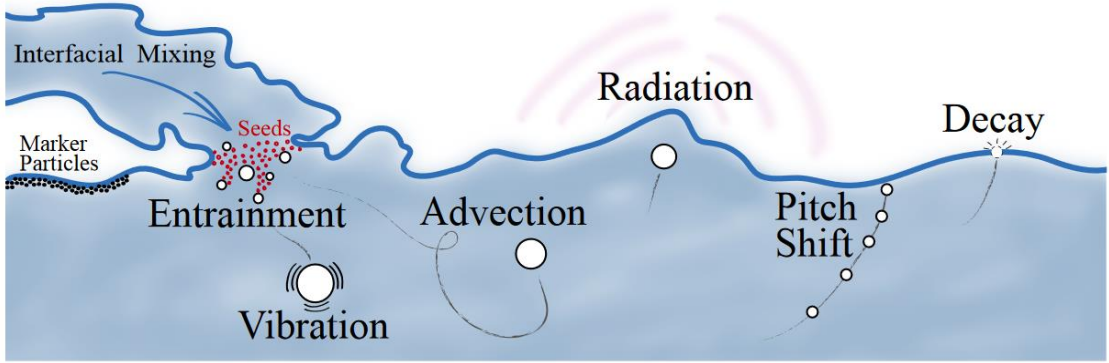

Figure 1　Life of an acoustic bubble (Zheng and James. 2009)

In 1933, Minnaert established a model to explain the principle of sound generation through bubble pulsation and calculated the resonant frequency (Minnaert, 1933). The results indicate that the frequencies of the bubbles follow a simple equation.

$$f_M = \frac{1}{2\pi r}\sqrt{\frac{3\kappa P_0}{\rho}}$$

In the above equation, $f_M$ is the Minnaert frequency of the bubble, $r$ is the equilibrium radius of the bubble, $\kappa$ is the adiabatic index of the gas within the bubble, $P_0$ is the standard pressure, and $\rho$ is the density of the liquid.

For a single bubble in water at standard pressure, the above equation reduces to

$$f_M = \frac{1}{2\pi r}\sqrt{\frac{3 \times 1.400 \times 100000}{1000}} = \frac{3.26(m/s)}{r}$$

It is amazing that the frequency of the bubble is simply inversely proportional to the radius. With the deeper study of bubble dynamics, the Rayleigh-Plesset-Noltingk-Neppiras-Poritsky (RPNNP) equation has been developed by taking into account surface tension, vapor pressure, and viscosity. A more generalized formula for bubble frequency can be written as follows (Nelli et al., 2022).

$$f_M = \frac{1}{2\pi r\sqrt{\rho}}\sqrt{3\kappa\left(P_0 - p_v + \frac{2\sigma}{r}\right) - \frac{2\sigma}{r} + p_v - \frac{4\mu^2}{\rho r^2}}$$

Here $p_v$ is the vapor pressure, $\sigma$ is the surface tension, and $\mu$ is the viscosity. During pouring, entrained bubbles have complex shape and size distributions, resulting in complex acoustic characteristics. The amplitude and spectral characteristics of the sound are shaped by the collective dynamics of the bubble cloud and the nonlinear interactions between the bubbles and the acoustic field.

## 2.4 Identifying the three sounds on the frequency spectrum

In order to verify the presence of the above three sound sources, we recorded the sound of pouring water and analyzed its spectrum. The spectrum analysis method, using STFT (short-time Fourier transform) to transform a series of audio signal from time vs amplitude to time vs frequency, which is widely used in audio analyzing because the researcher can view a signal in a frequency axis. Due to the poor experimental conditions, we could only record in a normal room. A beer tower purchased from Taobao is utilized to produce a steady water flow. A Re-de-kuai (a traditional Chinese appliance for heating water) purchased from Taobao is used to heat the water inside. The temperature of cold and hot water is 10 ℃ and 90 ℃, respectively. An EarthWorks M23R measurement microphone which has nearly pure-flat spectral response in common air temperature range, is used to record the sound of pouring water for better accuracy. The overall setup is shown in Figure 2 (poor but effective!).

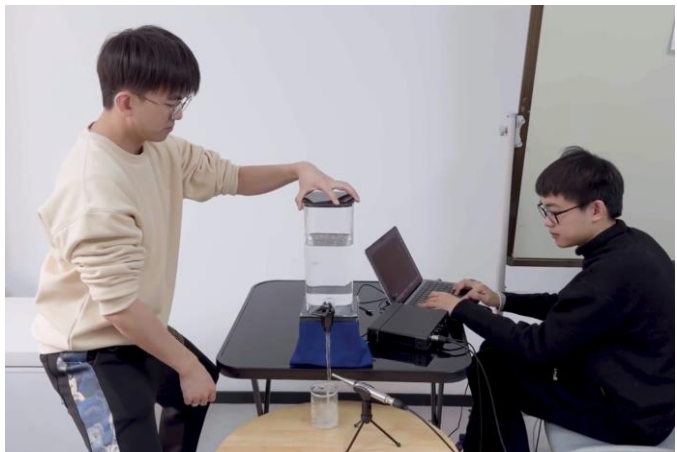

Figure 2 The experimental setup of the recording process

Figure 3 illustrates the spectrogram of the sound of pouring hot and cold water. A rising band and a falling band can be identified, which correspond to the air resonance and the vibration of the beaker and water, respectively. Specifically, the frequency of the air column resonance gradually rises from the initial 700 Hz to 2000 Hz, approximately. And the vibration of the beaker and water goes down from 7000 Hz to 200 Hz, approximately.

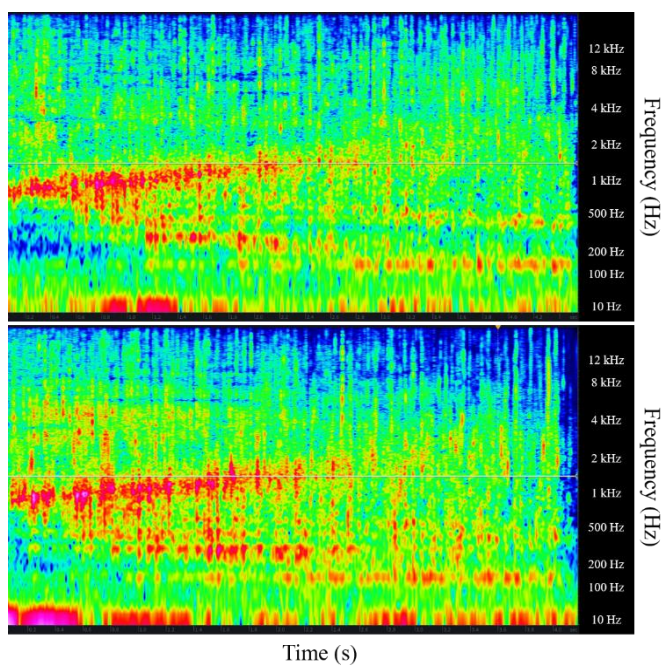

Figure 3  Spectrogram of audio samples of pouring cold (top) and hot (bottom) water. (The transverse line stands for 1500 Hz.)

Apart from the two seemingly continuous frequency bands, numerous transient sounds are distributed across the frequency domain. Notably, a significant number of transient sounds can be observed above 1000 Hz, which we postulate corresponds to bubble sounds. Grounded in empirical observations, we recognize that the bubble size typically ranges in the millimeter to centimeter scale, resulting in corresponding Minnaert resonance frequencies within the range of hundreds to thousands Hz. These bubble sounds are transitory in nature. The rapid, simultaneous generation of a large population of bubbles exhibiting a distribution of sizes gives rise to an ensemble of transient sounds that collectively span the entire audible frequency spectrum.

## 3. Which sound causes the essential acoustic difference between hot and cold water?

The phenomenon of hot and cold water sounding different must be attributable to the influence of temperature on the frequency characteristics of the constituent sound sources through an underlying physical mechanism. It is therefore imperative to critically examine which of the three principal sound components contributes most profoundly to the perceived frequency disparities.

(1) The air in the vessel is heated when pouring hot water. The rise in the speed of sound leads to a rise in the air resonance frequency. The overall upward shift of the band of air resonance is visually apparent in Figure 3. Since the sound speed in air is 337.4 m/s at 15 °C and 381.7m/s at 90 °C, we would expect a maximum of 13% increase in the tone of air resonance in theory. This increase is not large, and in actuality the air temperature can't possibly reach 90°C. Thinking back to life experience, most people do not accurately distinguish the presence of air resonance, let alone perceive small changes in it.

(2) Temperatures varying in the range of 10 °C to 90 °C have a minor effect on the physical properties of glass. Thus, temperature does not have an obvious effect on the frequency of glass and water vibrations.

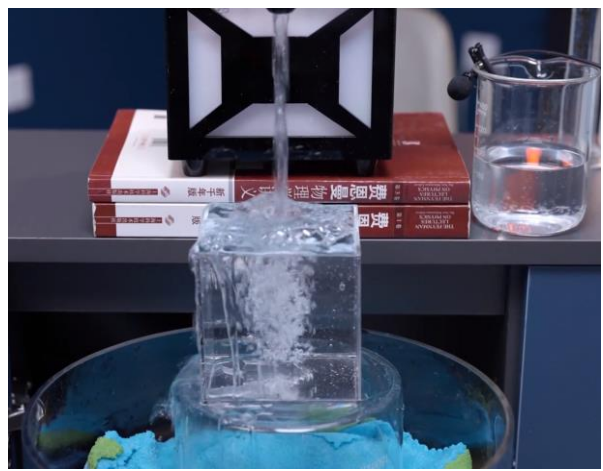

Figure 4 Experimental setup to capture the bubble sound

Intuition tells us that the bubble sound contributes the most significant difference. To experimentally assess this conjecture, we modified the experimental setup by initially filling the vessel and subsequently allowing the water to overflow, as depicted in Figure 4. Many towels were placed in the container below to prevent the overflowing water from producing sound (silly looking but effective!).

At this point since there isn't any air in the container, there is no air resonance sound. Furthermore, the water level remains consistently full throughout the process, resulting in a constant frequency response arising from the vibrations of the vessel and the water. Consequently, the dominant sound is the steady, isolated sound from the bubbles produced by the impinging water jet upon entering the bulk fluid phase.

The spectrogram of pouring hot and cold water into water-filled vessels are shown in Figure 5. The frequencies were quite stable during the 30 seconds of the pouring process, which also confirms the previous analysis of the 3 sound sources. The points arranged in the vertical direction in Figure 5 may represent the transient sound that is emitted when large amounts of bubbles are formed instantaneously.

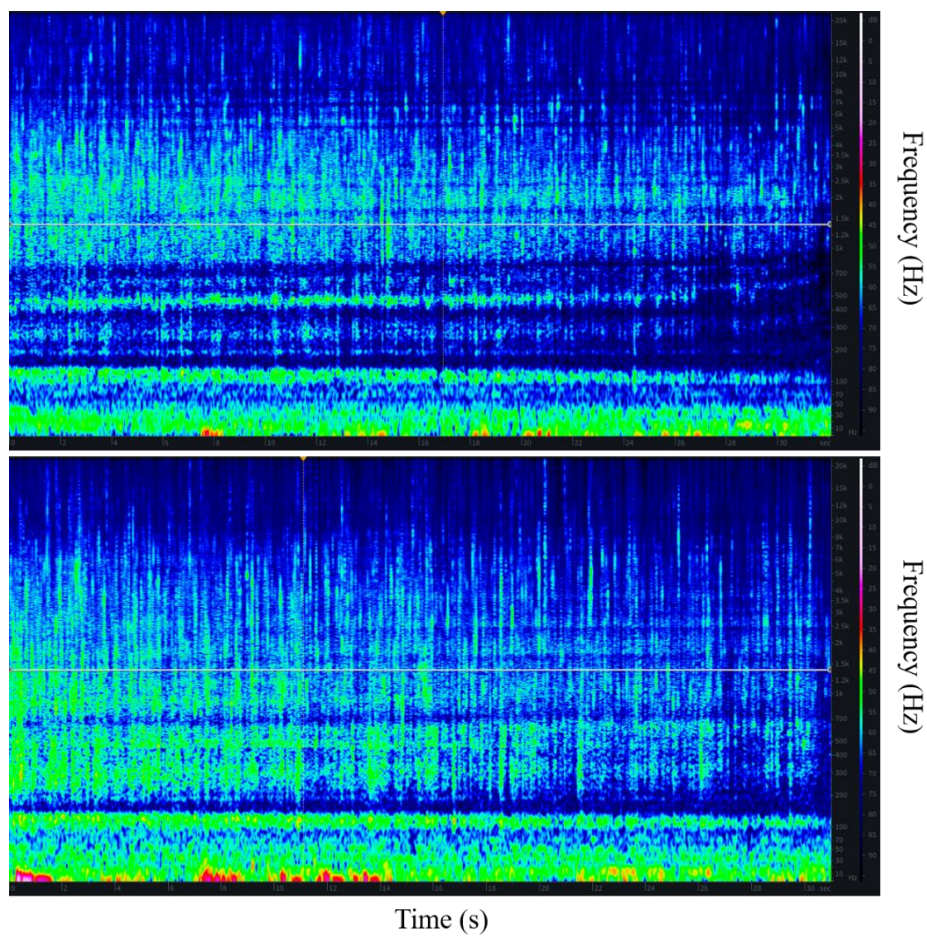

Figure 5  Spectrogram of audio samples of pouring cold (top) and hot (bottom) water into filled vessels

Two representative audio recordings of pouring cold and hot water are attached in the Supplementary Material for the readers. We believe that by virtue of excluding the sounds of air, vessel, and water, one can still clearly discern auditory distinctions between cold and hot water samples. From this observation, it can be reasonably inferred that the sound produced by the entrained bubbles encapsulates the most essential features of the sounds of pouring water.

Of course, the authors must admit that this conclusion drawn represents a subjective extrapolation derived from the personal experiences and perceptions of the authors and the individuals that the authors interviewed. A rigorous argument for this postulate may require the design and execution of methodical blind testing experiments grounded in the principles of sensory science. But we believe that once you listen to the audio in the Supplementary Material, you will be delighted to experience an intuitive moment: Ah! Funny! That's the sound!

## 4. How is the acoustics of pouring hot and cold water different?

In order to investigate the key distinction between the acoustic properties of hot and cold water, we plotted their energy spectra based on the entire recoding samples , then add a 1/6 octaves smoothing for better resonating human ears characteristics, as shown in Figure 6. Since our objective is to ascertain the relative intensities across the frequency domain rather than the absolute magnitudes of the acoustic energy, we shall focus our analysis on examining the distribution profiles of the relative spectral energy content for the hot and cold water cases as presented in Figure 6.

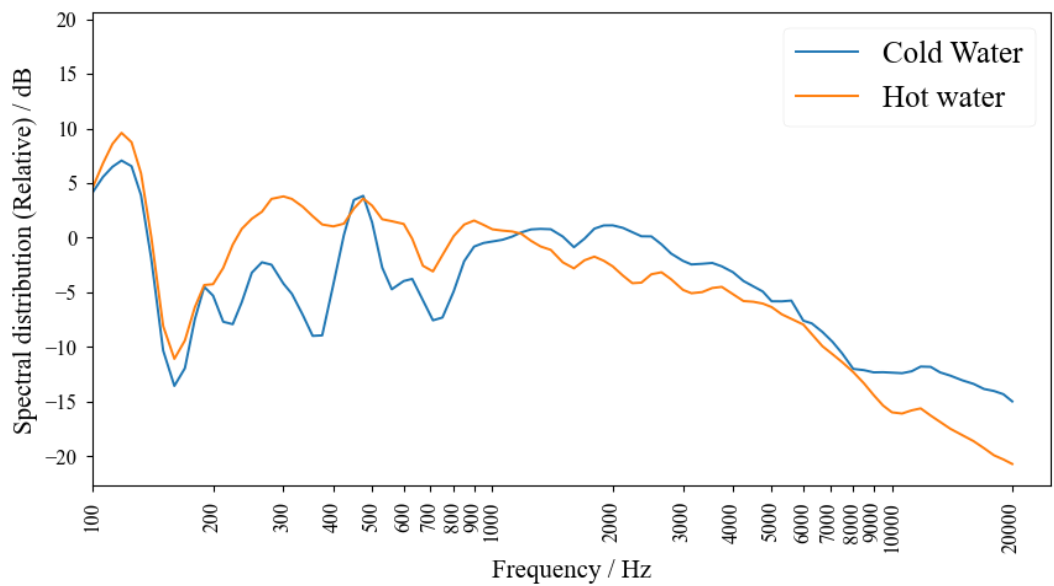

Figure 6   The spectral energy distributions of pouring hot and cold water sampled from whole recording, 1/6 octaves smoothing.

The spectral energy distributions clearly illustrate that the sound of pouring hot water exhibits a predominance of energy concentrated in the lower frequencies, while that of cold water displays a predominance of energy at higher frequencies within the audible range. Stated alternatively, the sound of hot water contains more low-frequency components, whereas the sound of cold water contains more high-frequency components. This is why, intuitively, even a person who doesn't know anything about acoustics would qualitatively describe hot water as sounding duller and cold water as sounding brighter or crisper.

## 5. Why does pouring hot water produce more low frequency sound?

To bridge the acoustics of pouring water with its underlying physics, we utilized a camera operating at 800 frames per second (fps) to observe the size of bubbles generated by pouring both cold and hot water. Figure 7 presents some typical comparisons of these observations.

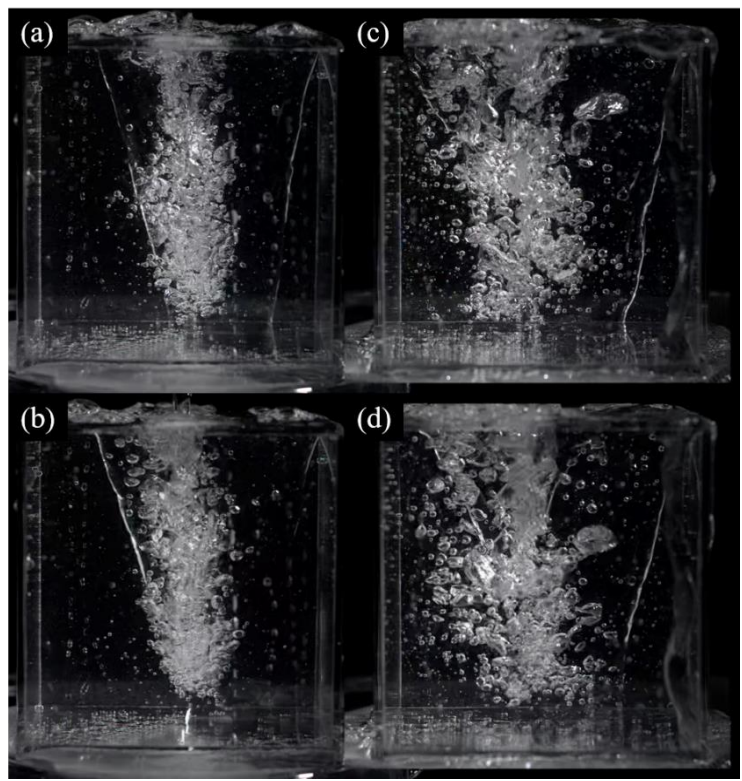

Figure 7    Comparison of bubble size in cold (a and b) and hot water (c and d).

Overall, pouring cold water tends to generate a higher quantity of smaller bubbles with radii ranging from 1 to 2 millimeters, whereas pouring hot water can produce larger bubbles with radii of 5 to 10 millimeters. However, it is important to note that these are merely qualitative observations made by visual inspection. To arrive at accurate conclusions, a quantitative analysis of the bubble size distribution is necessary. We look forward to more in-depth research by readers interested in this topic.

As previously mentioned, the Minnaert resonance law stipulates that the frequency is inversely proportional to the radius of the bubble. Under usual conditions, the formula can be roughly simplified to $f = 3/r$. Consistent with the Minnaert resonance law, bubbles with a radius of 1-2 millimeters emit frequencies approximately ranging from

1500-3000 Hz, whereas bubbles with a radius of 5-10 millimeters produce frequencies around 300-600 Hz. It is amazing that these findings align with the pattern observed in Figure 6, where hot water exhibits a richer composition at the $10^2$ Hz frequency level, while cold water has a richer composition at the $10^3$ Hz frequency level. We were surprised to discover that the law of Minnaert resonance for bubbles serves as the precise bridge between acoustics and fluid mechanics that we had been seeking.

In summary, we provide the conjecture that temperature affects the pouring sound by changing the size of the bubbles produced. Specifically, pouring hot water leads to the generation of a greater number of large bubbles compared to pouring cold water. Furthermore, these larger bubbles emit more low-frequency sound at the instant of their formation when they undergo the Minnaert resonance, resulting in an audible difference between hot and cold water.

## 6. Why does pouring hot water produce larger bubbles?

The theoretical study of bubble entrainment processes has been quite extensive (Biń, 1993; Brouilliot and Lubin, 2013; Endoh, 1982; Hwang et al., 1991; Zhu et al., 2000), but the complexity of hydrodynamics has left the effect of temperature on the nature of bubble entrainment yet to be fully investigated.

Callaghan et al. studied bubbles entrainment caused by water jet at temperatures ranging from 5 °C to 30 °C, and found that higher temperatures lead to the creation of larger bubbles (Callaghan et al., 2014). They attributed this phenomenon to the decreased viscosity of hot water, which led to a higher Reynolds number and consequently increased the roughness of the jet surface. Recent studies have also demonstrated that the roughness or steepness of the jet surface is the primary factor responsible for generating the pouring sound (Boudina et al., 2023). Chen and Guo observed that as the viscosity of the liquid increases, the size of the bubbles generated by the impact of droplets into a pool decreases (Chen and Guo, 2014). The influence of temperature on the size of bubbles produced by the jet may be attributed to various factors such as the viscosity of the liquid, surface tension, and vapor pressure.

We believe that the most likely explanation is the increase in turbulence intensity due to the low viscosity. To test this conjecture, some rudimentary experiments can be conducted, such as observing whether bubbles formed from jets of hot sugar water, which has a viscosity comparable to cold water, are smaller than those produced in hot water. However, a definitive conclusion will have to await further analysis by fluid mechanics experts.

## 7. Conclusion

The fascinating phenomenon that people can hear the difference between the sound of pouring hot water and pouring cold water has garnered widespread attention on social media. While this observation may appear trivial at first glance, a closer examination reveals that it is rooted in a multifaceted scientific mechanism, leaving us in awe of the wonders of the science behind it. After some rudimentary experimentation by amateur science enthusiasts like us, we propose a plausible explanation on the observed phenomenon.

(1) During the pouring process, the free-falling water jet generates bubble entrainment within the water in the vessel.

(2) When pouring hot water, it produces a higher quantity of large bubbles compared to pouring cold water. This phenomenon can be attributed to the decreased viscosity of hot water, which leads to a greater degree of turbulence and consequently a rougher surface of the jet.

(3) At the instant of their creation, the bubbles undergo pulsation and become audible due to the Minnaert resonance. Larger bubbles emit sound at a lower frequency.

(4) The abundance of larger bubbles in hot water generates a higher proportion of low-frequency sound components, whereas the smaller bubbles in cold water produce more high-frequency sound components. This accounts for the audible difference between the sound of pouring hot water and cold water.

There are indeed limitations in this study, including the lack of rigorous experimentation and the subjective nature of some extrapolations. The authors have neither rich professional knowledge nor professional experimental equipment. Actually, the author has nothing to do with the subject, except a curious mind and a desire to learn. It is understandable that some readers may view this article as civilian science after reading it.

Nevertheless, we sincerely and humbly expect that this article can serve as a small spark of inspiration for readers who are curious about this issue. We further hope that the problem of the acoustic distinction between hot and cold water will finally be solved. If you find that this paper does offer you some novel ideas, we would be delighted. Conversely, if you identify any errors in this article, we implore you not to laugh at the authors' amateurish endeavors.

# Acknowledgement

The authors would like to thank Xiyang Liu for helping with the high-speed photography. Special thanks to Sophia Zhang for her strong support as always.